**Full title**

Ecosystem models cannot predict the consequences of conservation decisions

**Short title**

Poor predictive power of ecosystem models

**Authors**

Larissa Lubiana Botelho[1,2,3*], Cailan Jeynes Smith[1], Sarah Vollert[1,3], Michael Bode[1,2]

**Affiliations**

1. School of Mathematical Sciences, Queensland University of Technology, Brisbane, Australia
2. Securing Antarctica's Environmental Future, School of Mathematical Sciences, Queensland University of Technology, Brisbane, 4000, Australia.
3. Centre for Data Science, Queensland University of Technology, Brisbane, Australia

**\* Contact author**

Larissa Lubiana Botelho

School of Mathematical Sciences,

Queensland University of Technology

4 George St, Brisbane 4000, Australia

Email: larissa.lubianabotelho@hdr.qut.edu.au

**SIGNIFICANCE STATEMENT**

Applied ecological modelling aims to predict ecosystem responses to management actions, to enhance resource allocation and prevent unintended consequences. However, ecosystems are complex, uncertain, and nonlinear, which has sparked a debate about whether ecosystem models can provide reliable forecasts of future dynamics. We use experimental data from microcosms to calibrate ecosystem models and test their predictive power. The results are confronting: even with accurate timeseries data on the abundance of all species in the ecosystem, our calibrated models were unable to provide insights into the ecosystem dynamics, make accurate short-term forecasts, or choose an effective management intervention. The inability of these models to handle simple microcosms casts doubts on their ability to support management decisions in complex, real-world ecosystems.

**ABSTRACT**

Ecosystem models are often used to predict the consequences of management decisions in applied ecology, including fisheries management and threatened species conservation. These models are high-dimensional, parameter-rich, and nonlinear, yet limited data is available to calibrate them, and they are rarely tested or validated. Consequently, the accuracy of their forecasts, and their utility as decision-support tools is a matter of debate. In this paper, we calibrate ecosystem models to time-series data from 110 different experimental microcosm ecosystems, each containing between three and five interacting species. We then assess how often these calibrated models offer accurate and useful predictions about how the ecosystem will respond to a set of standard management interventions. Our results show that for each timeseries dataset, a large number of very different parameter sets offer equivalent, good fits. However, these calibrated ecosystem models have poor predictive accuracy when forecasting future dynamics and offer ambiguous predictions about how species in the ecosystem will respond to management interventions. Closer inspection reveals that the ecosystem models fail because calibration cannot determine the types of interactions that occur within the ecosystem. Our findings call into question claims that ecosystem modelling can support applied ecological decision-making when they are calibrated against real-world datasets.

**INTRODUCTION**

A primary goal of applied ecological modelling is to provide accurate and actionable predictions about how ecosystems will respond to management actions [1,2]. Accurate predictions can help managers make the most of limited resources, and avoid unintended outcomes [3,4]. The interaction between different species in an ecosystem often determines the consequences of management interventions, and so ecosystem models are frequently used to support conservation management decisions in two ways. First, these models can be used to provide quantitative predictions of the future ecosystem states [5,6]. Second, the structure of these models can provide a mechanistic explanation for the ecosystem dynamics, offering causal explanations for positive and negative outcomes [2]. In these ways, ecosystem models have been used to support marine and terrestrial invasive species management [7–9], threatened species reintroductions [10], understand climate change and cumulative impacts [11,12], make fishery management decisions [13,14], and guide the introduction of novel biocontrol agents [15].

However, ecosystem models face several challenges as predictive tools. Ecosystems are complex, containing a large number of components (e.g., species) that are connected by a dense web of nonlinear interactions. Describing and predicting the dynamics of such systems is difficult [16–18], to put it mildly. Even for short-term forecasts, ecosystem models need a very precise understanding of the model structure [19], parameter values [20–22], and initial conditions [23]. Considerable amounts of data are required to estimate these model components. Ecosystem models are often calibrated using time-series abundance data of a subset of species in a community [6,9,21,24–28]. However, these data are generally noisy, short, sparsely sampled, and unreplicated [4,6,9,29].

Claims that ecosystem models can offer useful predictions for conservation management are often based on *in silico* analyses, where models are calibrated using simulated and noisy time-series data [5,6,30]. This is because *in situ* data are expensive, challenging to collect, and impossible to replicate. However, the results of *in silico* analyses can overstate the predictive power of ecosystem models. By calibrating the synthetic data with the same model that was used to generate it, these analyses ignore the possibility of structural uncertainty (i.e., model misspecification) [31]. A more rigorous test of ecosystem models would require them to be calibrated using data from real ecosystems.

In this paper, we use ecological microcosm experiments to provide the diversity and amount of real data that is required for a rigorous *ex silico* test of ecosystem models. Our findings are pessimistic. Firstly, the calibrated models produce inaccurate forecasts, compared to simple null models. Secondly, We find that the calibrated ecosystem models give ambiguous predictions about how ecosystems will respond to management actions. Finally, we find that this lack of predictive power stems from confusion about the structure of the ecosystem interactions. As a consequence, our results suggest that caution is necessary when using ecosystem models as conservation decision-support tools.

**MATERIALS AND METHODS**

**Overview**

Ecosystem models are often used as conservation decision-support tool for conservation actions to forecast the consequences of altering the abundance of a "perturbed" species on other species in the ecosystem. For example, managers might reduce the abundance of an invasive predatory species on an island, and use an ecosystem model to predict the benefits for prey species of conservation concern, or for other invasive species [32,33]. As another example, fishery managers might consider increasing the allowable catch for a commercially valuable species, and use an ecosystem model to predict the consequences for other species which either prey upon, or are predated upon, by the fished species [34].

To predict the response of other species in the ecosystem, a manager would begin by measuring the abundance of many species in the ecosystem over a period of time. They would then use this abundance time-series data to calibrate an ecosystem model, comprising a set of coupled differential equations that represent each of the observed species. Finally, they would use the calibrated ecosystem model to predict the response of important species in the ecosystem to a novel press perturbation that affects the perturbed species. This is the basic approach used by previous evaluations of ecosystem models [5,6,9,35], and we approach the data from the microcosm experiments as we expect an applied ecologist would.

**Experimental microcosm time-series data**

Pennekamp and colleagues [36] created and shared a dataset which contains time-series from multiple microcosm ecosystems. Each dataset reports the abundance of up to five interacting unicellular aquatic ciliate species. The microcosm ecosystems were run for 19 days under controlled conditions, and the abundance of each species was estimated daily using video sampling. While this produces an measure of abundance that is more accurate than most field ecology studies, the data still include some amount of demographic stochasticity and sampling error.

From the full set of experiments, we omit datasets that contained the species *Tetrahymena thermophila*, which rapidly declined under all conditions, and can therefore not be considered a viable species

(following ref [27]). We also excluded datasets with fewer than 3 species, to focus on ecosystems with more complex dynamics. There were 110 different microcosms remaining.

**Calibrating the ecosystem models**

We follow standard practice [9,26,27,36,37] by fitting generalised Lotka-Volterra models to the time-series data:

$$\frac{1}{n_i}\frac{dn_i}{dt} = r_i + \sum_{j=1}^{S_e} a_{ij}n_j + p_i(t)$$

Equation 1

where $S_e$ is the total number of species in experiment $e$, and $n_i$ is the modelled abundance of species $i$. To calibrate this model, we need to estimate a set of parameters $P$, containing the intrinsic growth rate of each species $r_i$; the per capita influence of species $j$ on species $i$, $a_{ij}$; and the initial abundance of each species, $n_i(t = 0)$. The final term on the right-hand side, $p_i(t)$, is used to simulate the effects of a conservation action (see below).

Following Rosenbaum & Fronhofer [26], we identify parameter sets that provide close fits to the microcosm time-series data, using the sum of squared deviations (SSD) between the observations and the predictions. Because each species has vastly different average abundances, we standardise each species' abundance by its average value to avoid the more abundant species dominating the fit statistic.

Given the high dimensionality of the model (e.g., 15 parameters for 3 species, and 35 parameters for 5 species), we expect that more than one set of parameters will offer a good-fit to the data. Visually, this is certainly the case (Figure 1). Choosing any single one of these parameters sets as the "best-fit", simply because its SSD is slightly lower than the others, would put too much emphasis on our choice of fit statistic, our fitting algorithm, and our dataset. We therefore repeatedly apply a gradient descent search from multiple initial conditions and accept any local minima that provide an "equivalently" good fit to the data, which we define as parameter sets with the top 10% of SSD values (*Supplementary Figure S1*).

We continue the search until we identify 500 calibrated ecosystem models for each of the 110 microcosm experiments. See *Supporting Information* for more detail on the fitting process.

**Validating the calibrated ecosystem models**

To inform environmental management decisions, calibrated models should accurately forecast each species' abundance, at least in the near future. To test prediction accuracy, we therefore calibrate the ecosystem models using the first 18 data-points in the time-series, and then use these parameterised models to forecast the abundances of each species at the end of the experiment. Specifically, we tested whether any of the calibrated models could forecast the $t = 19$ abundance of all species, within $\pm 25\%$ of their true value.

We also assessed the accuracy of the ecosystem models by comparing them to the predictions of a simple null model, where we assume that each species abundance will not change in the final timestep, i.e., $n_i(t = 19) = n_i(t = 18)$ (see Supplementary Material Section 2.2). To justify their use, the calibrated ecosystem models should be able to easily generate more accurate forecasts than this evidently false null model.

**Predicting responses to conservation action**

Decision makers are mostly interested in predicting how particular species in an ecosystem will respond to planned interventions, compared to a counterfactual of non-intervention. On Australia's Hermite Island, for example, conservation managers planned to eradicate invasive cats, but wanted to predict how this would affect the abundance of invasive rats, two native marsupial species, and one native bird species [38]. On California's Channel Islands, managers needed to predict how native foxes and golden eagles would respond to the eradication of invasive pigs [39].

Faced with the type of data found in the microcosm datasets, we assume that a conservation decision-maker would first identify a set of ecosystem models that provide good fits to the observed time-series. They would then use this set of calibrated models to simulate a planned intervention, by perturbing one of the species in the system (e.g., controlling an invasive species). Finally, they would assess whether the planned perturbation would increase or decrease the abundance of species of conservation interest in the system. To be useful in this process, the set of calibrated ecosystem models would need to provide relatively unambiguous predictions [6,40]. For example, if the overwhelming majority of the models agreed that controlling the perturbed species would not result in a decrease in any of the other species in the system. By contrast, if half of the models predicted an increase, while the other half predicted a decrease, then the ecosystem models would not be useful.

We simulate interventions using a negative press perturbation on a single species in the ecosystem model by increasing the value of $p_i(t)$ in Equation 1 from zero to a non-zero value, after the end of the observed timeseries (when $t > 19$). For each of the 500 calibrated ecosystem models, we perturbed one species, and simulate the abundances of all species, both with and without this perturbation, for an additional 3 timesteps. We repeated this process for all species in all microcosms. We then assessed how often the models agree on whether the perturbation would increase or decrease the abundance of each species. When 85% or more of the set made the same prediction, we call the results of the fitting exercise "unambiguous".

**Understanding ecosystem interactions**

Ecosystem models also offer insights into functional relationships between species in the ecosystem. For each of our 500 calibrated ecosystem models, we classify the relationships between each pair of species as either mutualist ($a_{ij} > 0, a_{ji} > 0$), competitive ($a_{ij} < 0, a_{ji} < 0$), or predator-prey ($a_{ij} > 0, a_{ji} < 0$). To be useful, these insights also need to be unambiguous. For example, we would consider it useful if 95% of the calibrated ecosystem models agreed that species $i$ and species $j$ competed with one another. Across all the species interactions, in all the 110 experimental microcosms, we report the proportion of agreement shown across these calibrated models.

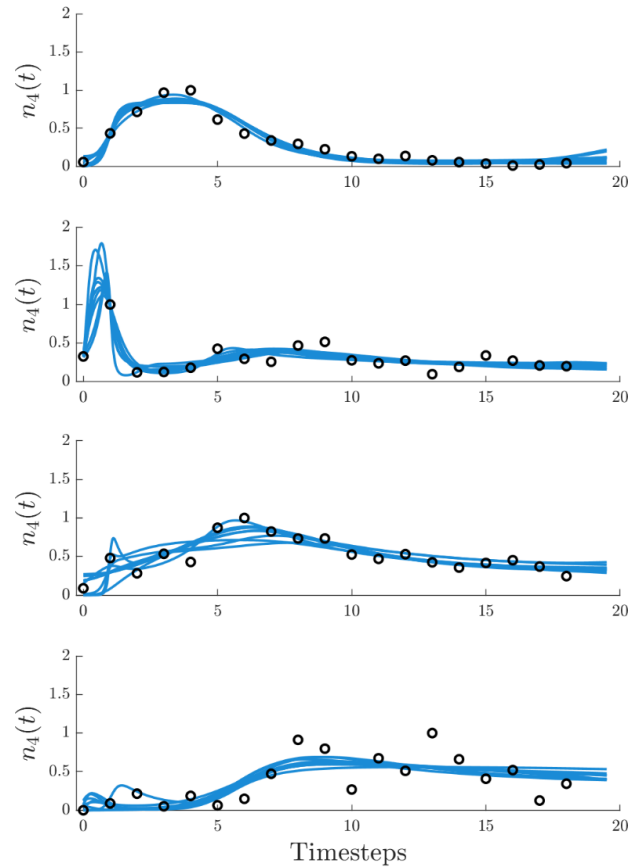

*Figure 1: Example fits to a single four-species microcosm ecosystem from the dataset. Blue lines show four different models that provide equivalently good fits to the observed time-series data (shown with black circles), according to our criterion.*

## RESULTS

**Model fitting**

The 500 calibrated models for each of the time-series provide visually equivalent fits (Figure 1) and produce comparable fit statistics (*Supplementary Figure S1*). The calibrated models capture the trends in each species' abundance, but do not attempt to match every abundance estimate. Calibrated models can contain either very similar, or very different, parameter estimates. Figure 2 shows an example of this: four different calibrated models (two in blue and two in red) which each provide visually and statistically good fits to a three-species microcosm dataset. The two blue parameter sets are very similar

(their Pearson's correlation coefficient is $\rho = 0.9$). By contrast, the parameter sets of the orange models are very different ($\rho = -0.25$).

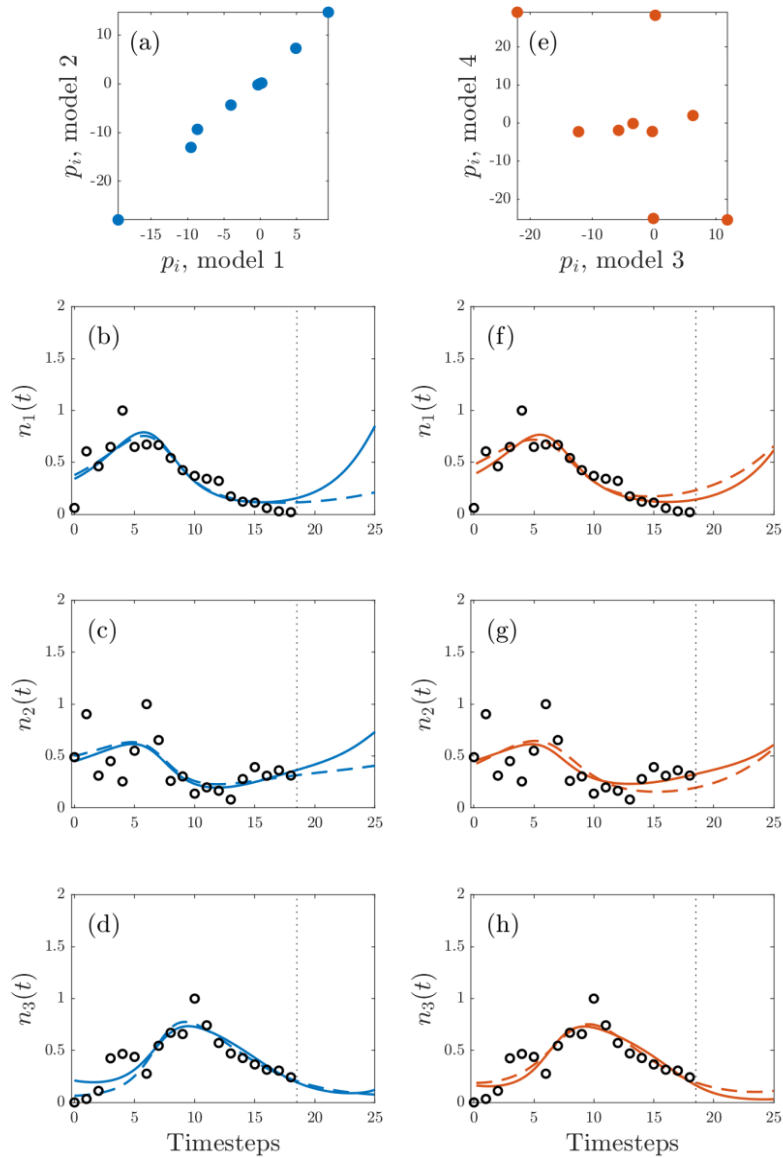

Figure 2: Comparison of four different calibrated ecosystem models for a 3-species microcosm dataset. All models provide equivalently good fits to the datasets. The two blue models have very similar parameter sets (panel a) but divergent forecasts (panels b-d). The two orange models have very different parameter sets, but congruent predictions (panels f-h).

The predictions made by these four different calibrated models are surprising. Despite their similar parameter sets, the two blue models make quite different predictions about species 1 and 2 – even after only 3 timesteps. By contrast, the two orange models make similar predictions, despite the dramatic differences in their parameter values.

**Ecosystem model validation**

While the ecosystem models provide close fits to the observed datasets, they perform poorly in validation analyses. We set our expectations for validation low – we assessed whether the calibrated ecosystem models could predict the abundance of all species within $\pm 25\%$ of their true values (*Supplementary Figure S3*).

Many models were able to predict the future abundance of one or more species to this level of accuracy, but across all microcosm datasets, considering the top 50 calibrated models, none were able to do so for all the species (*Supplementary Figure S4*).

**Predicting responses to conservation actions**

The calibrated models could also not agree on how species would respond to simulated conservation actions. Specifically, they could not reliably predict if other species in the system would be better or worse off, following a press perturbation to a single species, in comparison to a scenario in which no perturbation was performed.

Figure 3a shows the results for a randomly selected (but representative) five-species ecosystem. The models agreed on the effect of a negative press perturbation on a particular species would affect the abundance of that same species (i.e., the diagonal elements of the figure), but the predictions were ambiguous for almost all the other combinations (the off-diagonal elements). Figure 3b shows how often the calibrated models agreed on how any given species would respond to a negative press perturbation to another species in the ecosystem, across all 110 ecosystems, and for all combinations of perturbed species and response species. For 91% of simulated perturbations, the set of calibrated models can only return ambiguous predictions.

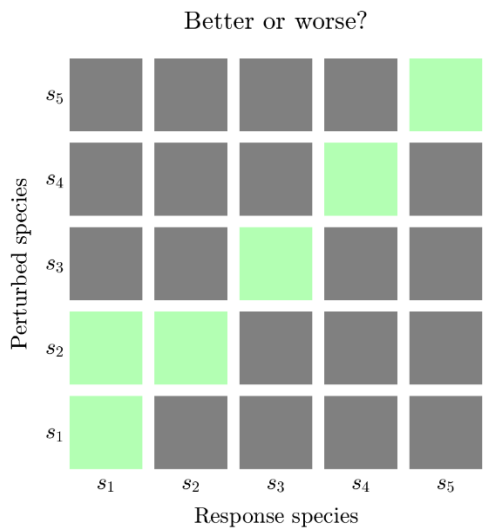 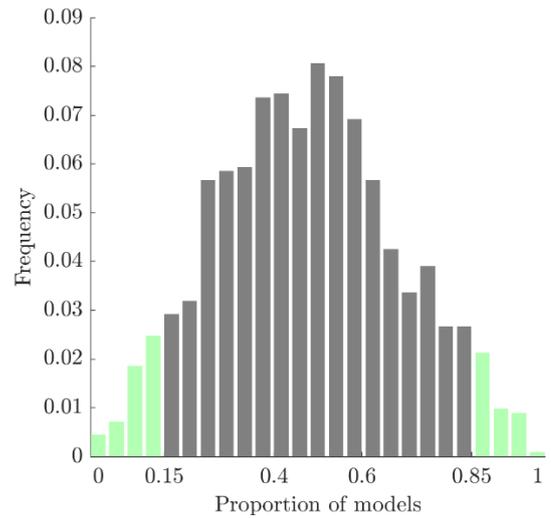

*Figure 3: Inconsistent predicted responses to a press perturbation. Panel (a) shows the consistency of responses to simulated press perturbations for a single exemplar 5-species experimental microcosm. A square is coloured green if 85% of the calibrated models agreed on how the column species would respond to a negative press perturbation to the row species. The square was coloured grey if the response was ambiguous. Panel (b) shows the same data, compiled across all 110 microcosm experiments. For 91% of simulated perturbations, the set of calibrated models could not make unambiguous predictions.*

**System description**

Finally, the different calibrated models do not agree on the character of interspecific relationships in the different microcosm ecosystems. Figure 4 shows that, in fewer than 1% of observed species interactions, the set of calibrated models unambiguously selected a particular type of interspecific relationship. Most interspecific interactions were ambiguously identified.

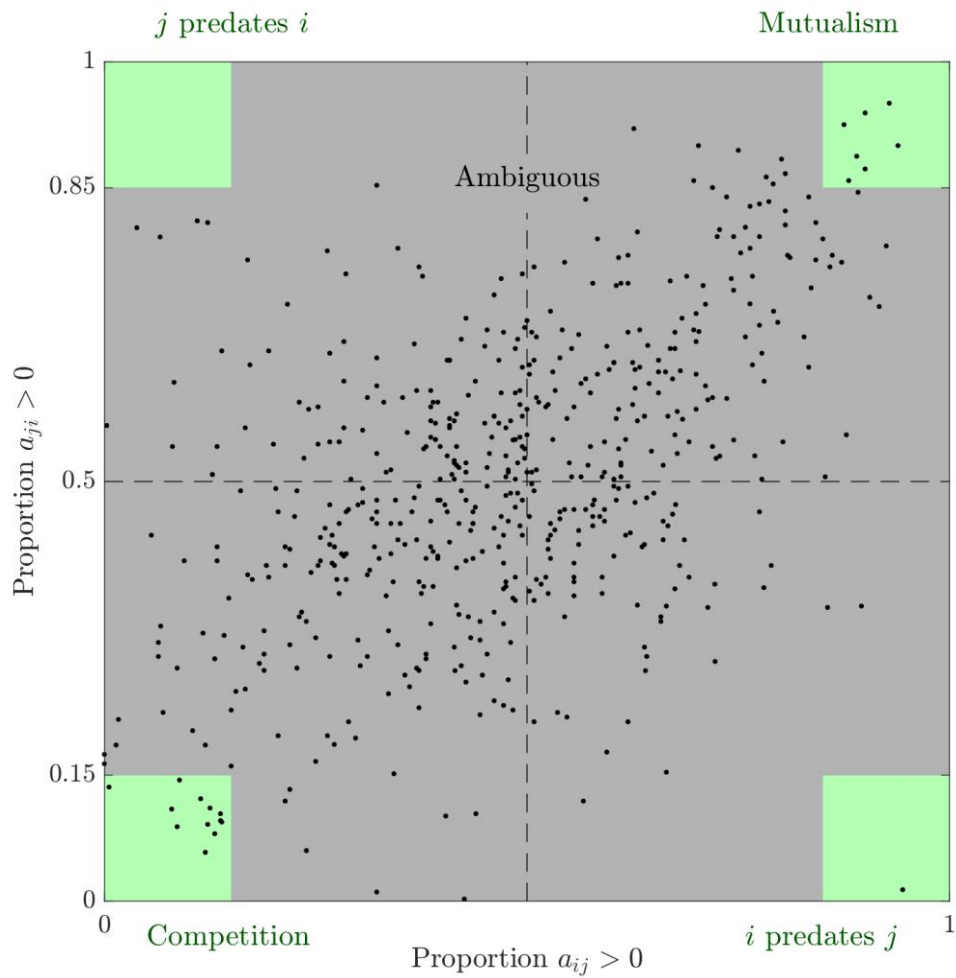

*Figure 4: Ability of the ecosystem models to identify the character of interspecific interactions. Each dot in this figure represents an interaction between two species in the microcosm dataset. The location of each dot indicates the proportion of the calibrated ecosystem models which agreed that the sign of $a_{ij}$ was positive. Green areas indicate >85% agreement. For example, dots in the lower left green square, 95% of the calibrated models agreed that the interaction was competitive (i.e., $a_{ij} < 0$ and $a_{ji} < 0$).*

**DISCUSSION**

Narrowly interpreted, our results demonstrate that we cannot predict the consequences of press perturbations to experimental microcosms or accurately describe the species interactions in these systems by fitting Lotka-Volterra models to abundance time-series. For each of the microcosm experiments, hundreds of calibrated models could accurately and equally replicate the observed time-series data. Despite their equivalent fits, these parameter sets (i) make inaccurate forecasts of future dynamics; (ii) disagree on how the ecosystems would respond to perturbations; and (iii) draw different conclusions about the nature of species interactions. Put simply, these ecosystem models – despite being calibrated to substantial datasets – have little value as tools for forecasting, decision-making, or understanding. Conservation managers face far more difficult questions than the ones we asked of our microcosm ecosystems. We did not ask, for example, how strong the control actions should be, when or where they should be applied, or whether additional interventions could mitigate any negative consequences. In a more general sense, we therefore believe that our results cast doubt on the utility of ecosystem models as tools for supporting decisions in real-world conservation settings.

Our conclusion that calibrated ecosystem models are uncertain is neither new nor particularly pessimistic [21]. However, our specific conclusions call into question some of the proposed solutions. In particular, ecosystem modellers are encouraged to undertake sensitivity analyses, to determine the robustness of their recommendations [41,42]. Local sensitivity analyses admit to being uncertain about the parameter values, but they assume that the true model lies somewhere in the vicinity of the best-estimate model. In contrast, our results show that numerous conflicting models offer equally accurate fits. The best-estimate model might therefore be entirely contradictory to the true ecosystem dynamics. Studies that apply local sensitivity analyses around these best-estimate parameter sets [43–46] will consequently be no more robust than ecosystem model analyses which base management recommendations on a single parameterisation [10,11,47], as they entirely ignore alternative ecosystem models which could plausibly generate the data. Bayesian methods could be used to capture the uncertainty in parameter values [48,49]. However, the magnitude of the uncertainty that we have encountered is more comparable to qualitative modelling approaches [20,33,50], which allow parameters to take any value, as long as the model has a particular sign structure. As with our findings (Figure 1), the predictions of such qualitative models are usually too ambiguous to support confident decision-making.

If we cannot make accurate or useful predictions about microcosm ecosystems, it is hard to believe that ecosystem models will have more success in real conservation contexts. It could be argued that our pessimistic conclusions are specific to our case-study: a microcosm ecosystem, fitted with deterministic, nonspatial Lotka-Volterra models, to a dataset with fewer than 20 noisy observations. Unfortunately, making predictions about real ecosystems is dramatically more difficult along each of these dimensions. Microcosms are small and closed, and their environments are kept constant and controlled. By contrast, real ecosystems contain hundreds of interacting species, spread across large, heterogeneous landscapes and seascapes, exposed to environmental variation and unpredictable disturbances. Thousands of parameters [51] are needed to capture these additional complexities, rather than the dozens of parameters required by microcosms. From a data perspective, our time-series are considerably superior than those generally available to conservation managers. Conservation datasets only estimate the abundance of a small fraction of the species in the system, and many of the observed "species" are in fact aggregates (e.g., "vegetation", or "invertebrates"; [52]). Abundances are rarely measured directly, and instead rely on proxies such as trapping rates [2], which are biased in confusing ways by season, environment, and observer.

Ecosystem models are attractive to conservation decision-makers because they offer a defensible, quantitative tool for making challenging choices, and because they are based on familiar ecological theory. However, our results reinforce a growing body of evidence which questions the utility of ecosystem models in applied ecology and conservation [33,51,53,54]. Ecosystem dynamics are too nonlinear, and ecosystem models are too flexible, for standard comparisons between models and data to offer predictive reliability [19,21]. Even when a large amount of calibration data is available, the set of models that can fit the data is too large, and too diverse, to offer actionable insights to decision-makers.

The underlying cause of this problem – nonlinearity and high-dimensionality – are intrinsic features of ecosystems. We therefore suspect that these are limitations of mechanistic ecosystem modelling in general, rather than the particular set of ecosystem models that are currently in use. Our recommendation is consequently that the field needs to approach predictive mechanistic models of species interactions with caution, since ecosystem modelling may face fundamental challenges. Nevertheless, decisions need to be made about ecosystem management. Whether through enhanced data collection practices, new developments in deterministic models or modelling to minimise risk rather than for accurate predictions - the field is in dire need of innovation. In comparably complex and uncertain decision-making fields, such as financial forecasting, modellers often eschew mechanistic approaches, despite acknowledging that the system dynamics are governed by mechanistic interactions [55]. Instead, decision-support models pay attention to phenomena such as covariation or autocorrelation, and focus on minimising risks and maintaining optionality. It is possible that such a change in focus – away from accurate predictions, towards risk minimisation and robustness – will deliver better decisions to conservation managers.

## ACKNOWLEDGEMENTS

Funding support was received from Australian Research Council Grant SR200100005 Securing Antarctica's Environmental Future.

Sarah Vollert acknowledges funding from an Australian Research Council Discovery Early Career Researcher Award (DE200100683) and a Queensland University of Technology Centre for Data Science, Australia Scholarship.

Cailan Jeynes-Smith is supported by an Australian Government Research Training Program Scholarship.

## REFERENCES CITED

1. Levins R. Discussion paper: the qualitative analysis of partially specified systems. Ann N Y Acad Sci. 1974;231: 123–138.
2. Lindenmayer DB, Wood J, MacGregor C, Foster C, Scheele B, Tulloch A, et al. Conservation conundrums and the challenges of managing unexplained declines of multiple species. Biol Conserv. 2018;221: 279–292.
3. McDonald-Madden E, Sabbadin R, Game ET, Baxter PWJ, Chades I, Possingham HP. Using food-web theory to conserve ecosystems. Nat Commun. 2016;7: 1–8.
4. Geary WL, Bode M, Doherty TS, Fulton EA, Nimmo DG, Tulloch AIT, et al. A guide to ecosystem models and their environmental applications. Nat Ecol Evol. 2020;4: 1459–1471.
5. Ives A, Dennis B, … KC-E, 2003 undefined. Estimating community stability and ecological interactions from time-series data. Wiley Online Library. 2003;72: 331–342.
6. Adams MP, Sisson SA, Helmstedt KJ, Baker CM, Holden MH, Plein M, et al. Informing management decisions for ecological networks, using dynamic models calibrated to noisy time-series data. Ecol Lett. 2020;23: 607–619.
7. Langseth BJ, Rogers M, Zhang H. Modeling species invasions in {Ecopath} with {Ecosim}: an evaluation using {Laurentian} {Great} {Lakes} models. Ecol Modell. 2012;247: 251–261.
8. Arias-González JE, González-Gándara C, Cabrera JL, Christensen V. Predicted impact of the invasive lionfish {Pterois} volitans on the food web of a {Caribbean} coral reef. Environ Res. 2011;111: 917–925.
9. Liao C, Xavier JB, Zhu Z. Enhanced inference of ecological networks by parameterizing ensembles of population dynamics models constrained with prior knowledge. BMC Ecol. 2020;20: 1–15.
10. Hunter DO, Britz T, Jones M, Letnic M. Reintroduction of Tasmanian devils to mainland Australia can restore top-down control in ecosystems where dingoes have been extirpated. Biol Conserv. 2015;191: 428–435. doi:10.1016/j.biocon.2015.07.030
11. Tulloch VJD, Plagányi ÉE, Matear R, Brown CJ, Richardson AJ. Ecosystem modelling to quantify the impact of historical whaling on {Southern} {Hemisphere} baleen whales. Fish and Fisheries. 2018;19: 117–137.

12. Bozec Y-M, Hock K, Mason R, Baird ME, Castro-Sanguino C, Condie SA, et al. Cumulative impacts across {Australia}'s {Great} {Barrier} {Reef}: {A} mechanistic evaluation. Ecol Monogr. 2021;91: 1–2.
13. Plagányi ÉE, Butterworth DS. A critical look at the potential of {Ecopath} with {Ecosim} to assist in practical fisheries management. Afr J Mar Sci. 2004;26: 261–287.
14. Marshall KN, Koehn LE, Levin PS, Essington TE, Jensen OP. Inclusion of ecosystem information in {US} fish stock assessments suggests progress toward ecosystem-based fisheries management. ICES Journal of Marine Science. 2019;76: 1–9.
15. Solé J, Estrada M, Garcia-Ladona E. Biological control of harmful algal blooms: {A} modelling study. Journal of Marine Systems. 2006;61: 165–179.
16. Beckage B, Gross LJ, Kauffman S. The limits to prediction in ecological systems. Ecosphere. 2011;2: 1–12.
17. May RM. Complexity and stability in model ecosystems. Princeton: Princeton UniversityPress. 1973.
18. Boit A, Martinez ND, Williams RJ, Gaedke U. Mechanistic theory and modelling of complex food-web dynamics in {Lake} {Constance}. Ecol Lett. 2012;15: 594–602.
19. Wood SN, Thomas MB. Super–sensitivity to structure in biological models. Proc R Soc Lond B Biol Sci. 1999;266: 565–570.
20. Baker C, Bode M, McCarthy M. Models that predict ecosystem impacts of reintroductions should consider uncertainty and distinguish between direct and indirect effects. Biol Conserv. 2016;196: 211–212.
21. Novak M, Wootton JT, Doak DF, Emmerson M, Estes JA, Tinker MT. Predicting community responses to perturbations in the face of imperfect knowledge and network complexity. Ecology. 2011;92: 836–846.
22. Aufderheide H, Rudolf L, Gross T, Lafferty KD. How to predict community responses to perturbations in the. Proceedings of the Royal Society B: Biological Sciences. 2013;280. doi:10.1098/rspb.2013.2355
23. May 1974 (Science - original bio-chaos paper). 2004; 1–4.
24. Schooler SS, Salau B, Julien MH, Ives AR. Alternative stable states explain unpredictable biological control of Salvinia molesta in Kakadu. Nature. 2012;469: 86–89. doi:10.1038/nature09735
25. Saade C, Kéfi S, Gougat-Barbera C, Rosenbaum B, Fronhofer EA. Spatial autocorrelation of local patch extinctions drives recovery dynamics in metacommunities. Proceedings of the Royal Society B: Biological Sciences. 2022;289. doi:10.1098/RSPB.2022.0543
26. Rosenbaum B, Fronhofer EA, Peters DPC. Confronting population models with experimental microcosm data: from trajectory matching to state-space models. Wiley Online Library. 2023;14: 4503. doi:10.1002/ecs2.4503
27. Maynard DS, Miller ZR, Allesina S. Predicting coexistence in experimental ecological communities. Nat Ecol Evol. 2020;4: 91–100.
28. Wootton KL, Curtsdotter A, Jonsson T, Banks HT, Bommarco R, Roslin T, et al. Beyond body size—new traits for new heights in trait-based modelling of predator-prey dynamics. PLoS One. 2022;17: e0251896. doi:10.1371/JOURNAL.PONE.0251896
29. White ER, Cox K, Melbourne BA, Hastings A. Success and failure of ecological management is highly variable in an experimental test. Proceedings of the National Academy of Sciences. 2019;116: 23169–23173.
30. Certain G, … FB-M in E and, 2018 undefined. How do MAR (1) models cope with hidden nonlinearities in ecological dynamics? Wiley Online Library. 2018;9: 1975–1995. doi:10.1111/2041-210X.13021
31. Hill SL, Watters GM, Punt AE, McAllister MK, Quéré C Le, Turner J. Model uncertainty in the ecosystem approach to fisheries. Fish and Fisheries. 2007;8: 315–336. doi:10.1111/J.1467-2979.2007.00257.X
32. Courchamp F, Langlais M, Sugihara G. Cats protecting birds: modelling the mesopredator release effect. Journal of Animal Ecology. 1999;68: 282–292.
33. Raymond B, McInnes J, Dambacher JM, Way S, Bergstrom DM. Qualitative modelling of invasive species eradication on subantarctic {Macquarie} {Island}. Journal of Applied Ecology. 2011;48: 181–191.
34. Harvey C, Cox S, … TE-IJ of, 2003 undefined. An ecosystem model of food web and fisheries interactions in the Baltic Sea. academic.oup.com.
35. Monsalve-Bravo GM, Lawson BAJ, Drovandi C, Burrage K, Brown KS, Baker CM, et al. Analysis of sloppiness in model simulations: Unveiling parameter uncertainty when mathematical models are fitted to data. Sci Adv. 2022;8. doi:10.1126/SCIADV.ABM5952/SUPPL_FILE/SCIADV.ABM5952_SM.PDF
36. Pennekamp F, Pontarp M, Tabi A, Altermatt F, Alther R, Choffat Y, et al. Biodiversity increases and decreases ecosystem stability. Nature. 2018;563: 109–112.
37. Fronhofer EA, Klecka J, Melián CJ, Altermatt F. Condition-dependent movement and dispersal in experimental metacommunities. Ecol Lett. 2015;18: 954–963.

38. Helmstedt KJ, Shaw JD, Bode M, Terauds A, Springer K, Robinson SA, et al. Prioritizing eradication actions on islands: it's not all or nothing. Journal of Applied Ecology. 2016;53: 733–741. doi:10.1111/1365-2664.12599
39. Courchamp F, Woodroffe R, Roemer G. Removing protected populations to save endangered species. Science (1979). 2003;302: 1532.
40. Peterson KA, Barnes MD, Jeynes-Smith C, Cowen S, Gibson L, Sims C, et al. Reconstructing lost ecosystems: A risk analysis framework for planning multispecies reintroductions under severe uncertainty. Journal of Applied Ecology. 2021;58: 2171–2184.
41. Hansen C, Drinkwater KF, Jähkel A, Fulton EA, Gorton R, Skern-Mauritzen M. Sensitivity of the Norwegian and Barents Sea Atlantis end-to-end ecosystem model to parameter perturbations of key species. PLoS One. 2019;14: e0210419. doi:10.1371/JOURNAL.PONE.0210419
42. Ayers MJ, Scharler UM. Use of sensitivity and comparative analyses in constructing plausible trophic mass-balance models of a data-limited marine ecosystem — The KwaZulu-Natal Bight, South Africa. Journal of Marine Systems. 2011;88: 298–311. doi:10.1016/J.JMARSYS.2011.05.006
43. White M, Thornton P, … SR-E, 2000 undefined. Parameterization and sensitivity analysis of the BIOME–BGC terrestrial ecosystem model: Net primary production controls. journals.ametsoc.org. [cited 26 May 2023]. Available: https://journals.ametsoc.org/view/journals/eint/4/3/1087-3562_2000_004_0003_pasaot_2.0.co_2.xml?tab_body=fulltext-display
44. Borer ET, Hosseini PR, Seabloom EW, Dobson AP. Pathogen-induced reversal of native dominance in a grassland community. Proc Natl Acad Sci U S A. 2007;104: 5473–5478. doi:10.1073/PNAS.0608573104
45. MacNeil M, Skiles J, Modelling JH-E, 1985 undefined. Sensitivity analysis of a general rangeland model. Elsevier. 1985;29: 57–76. Available: https://www.sciencedirect.com/science/article/pii/030438008590047X
46. Roemer GW, Donlan CJ, Courchamp F. Golden eagles, feral pigs, and insular carnivores: how exotic species turn native predators into prey. Proceedings of the National Academy of Sciences. 2002;99: 791–796.
47. Morello EB, Plagányi ÉE, Babcock RC, Sweatman H, Hillary R, Punt AE. Model to manage and reduce crown-of-thorns starfish outbreaks. Mar Ecol Prog Ser. 2014;512: 167–183.
48. Science MG-TC, 2008 undefined. Bayesian inference for differential equations. Elsevier. [cited 26 May 2023]. Available: https://www.sciencedirect.com/science/article/pii/S030439750800501X
49. Uusitalo L, Lehikoinen A, Helle I, & KM-EM, 2015 undefined. An overview of methods to evaluate uncertainty of deterministic models in decision support. Elsevier. [cited 26 May 2023]. Available: https://www.sciencedirect.com/science/article/pii/S1364815214002813
50. Peterson KA, Barnes MD, Jeynes-Smith C, Cowen S, Gibson L, Sims C, et al. Reconstructing lost ecosystems: {A} risk analysis framework for planning multispecies reintroductions under severe uncertainty. Journal of Applied Ecology. 2021.
51. Storch LS, Glaser SM, Ye H, Rosenberg AA. Stock assessment and end-to-end ecosystem models alter dynamics of fisheries data. PLoS One. 2017;12: e0171644.
52. Dexter N, Ramsey DSL, MacGregor C, Lindenmayer D. Predicting Ecosystem Wide Impacts of Wallaby Management Using a Fuzzy Cognitive Map. Ecosystems. 2012;15: 1363–1379. Available: http://link.springer.com/10.1007/s10021-012-9590-7
53. Bode M, Baker CM, Benshemesh J, Burnard T, Rumpff L, Hauser CE, et al. Revealing beliefs: using ensemble ecosystem modelling to extrapolate expert beliefs to novel ecological scenarios. Methods Ecol Evol. 2017;8: 1012–1021.
54. Baker CM, Bode M, Dexter N, Lindenmayer DB, Foster C, MacGregor C, et al. A novel approach to assessing the ecosystem-wide impacts of reintroductions. Ecological Applications. 2019. doi:10.1002/eap.1811
55. Haldane A. The Dog and the Frisbee. Federal Reserve Bank of Kansas City's 36th Economic Policy Symposium. Jackson Hole, Wyoming, USA: Federal Reserve Bank of the United States; 2012. p. 1.

# Supplementary Material to

## "Ecosystem models cannot predict the consequences of conservation decisions".

1. **Fitting the ecosystem models**

We fitted generalised Lotka-Volterra models to the time series-data

$$\frac{1}{n_i}\frac{dn_i}{dt} = r_i + \sum_{j=1}^{S_e} a_{ij} n_j + p_i(t)$$

where $n_i$ is the modelled abundance of species $i$, $r_i$ is the the intrinsic growth rate of species $i$, $a_{ij}$ is the per capita influence of species $j$ on species $i$, and the sum is performed over the total number of species, $S_e$, in the ecosystem $e$; $p_i(t)$ is the perturbation term.

We evaluate the model-data fit by comparing data simulated by the model (characterised by a choice of parameters **P**) to the experimental time-series data. We measure the fit, $D_e(\mathbf{P})$, of a particular parameter set using the sum of squared deviations (SSD) between the standardised observations

$$D_e(P) = \sum_{i=1}^{n}\sum_{k=1}^{19} |N_i^e(t_k) - n_i(t_k \vee P)|^2$$

We excluded any parameter sets that resulted in extinctions, since these were not observed in the dataset.

A search procedure must then be employed to find parameter values that minimise the SSD between the model and the observed data. We used a gradient-based, non-linear, multi-dimensional minimisation function (*Fmincon*; Matlab R2022) to search for the best-fit parameter set. This algorithm can find the local minimum of the SSD surface, $D_e(P)$, for a given initial condition. While the parameter combinations identified through this search will be local minima, we used multiple random initial search locations across the parameter space, which means that minima can be identified across a range of parameter values. We then only included fits that delivered the top 10% of SSD values (Figure S1)

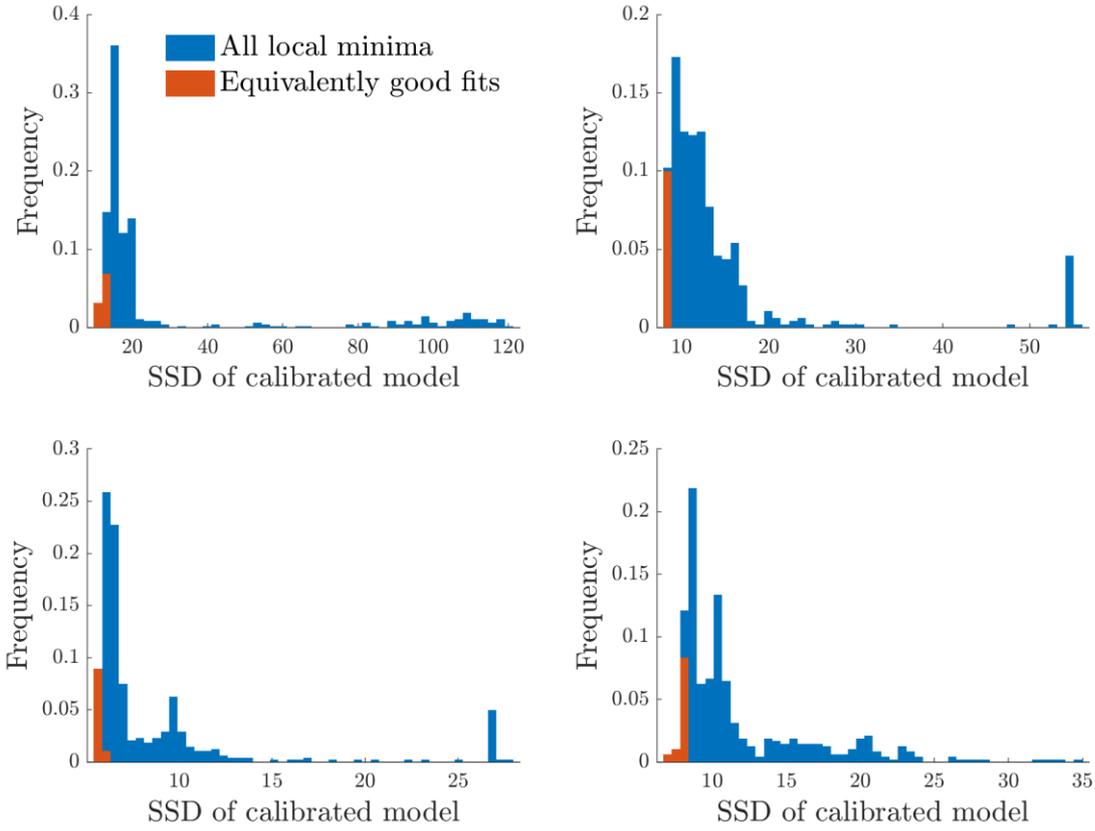

*Figure S1: Example fit statistics for the ecosystem model in Equation (1) to the experimental microcosm datasets. Blue histogram shows the SSD of a large number of local-minima model fits to the data. Red histogram shows the subset of these fits that we considered "equivalently good" fits.*

To measure the benefit of a conservation action for a species of interest by comparing the predicted future abundances with and without the intervention, we calculated:

$$m_P = \frac{n_i\left(t_p + T \vee p_j(t_p) = n_j(t_p)\right)}{n_i\left(t_p + T \vee \left(p_j(t_p) = 0\right)\right)}$$

Here, the numerator is the abundance of the perturbed species after a press perturbation has been performed at time $t_p$ on species $j$, and the denominator is the abundance of the species of interest at some time $t_p + T$ without any perturbation.

2. **Assessing predictive accuracy**

We assessed the ecosystem models' accuracy when forecasting populations using two complementary approaches. Firstly, we assessed their ability to predict populations within some margin of error of the truth.

Secondly, we compared their prediction error to the simplest possible model – a null model – where we predict no changes in populations through time.

In both analyses, we fit the ecosystem model to the first 18 datapoints many times and used these models to predict the populations at the 19th time-step, which we then compared to the observed populations in that timestep.

*2.1 Predictions within error margins*

The percentage relative error of the two models (the ecosystem models and the null model) for a single species can then be calculated for each prediction as

$$R_E = \frac{n_i^m - n_i^o}{n_i^o},$$

where $R_E$ is the relative error, $n_i^m$ is the modelled abundance of species $i$ in timestep 19, and $n_i^o$ is the observed population in timestep 19. Figure S2 shows these model fits and predictions for a 4-species microcosm experiment, where each model is coloured according to bounds on the relative error. In a conservation context, the accuracy needed from these predictions is problem-specific, however, we show three different bounds to indicate a variety of different accuracy targets. Note that when populations are close to zero, less absolute error in predictions is tolerated, however, this is often true in the conservation context where accuracy in small or threatened populations is critical.

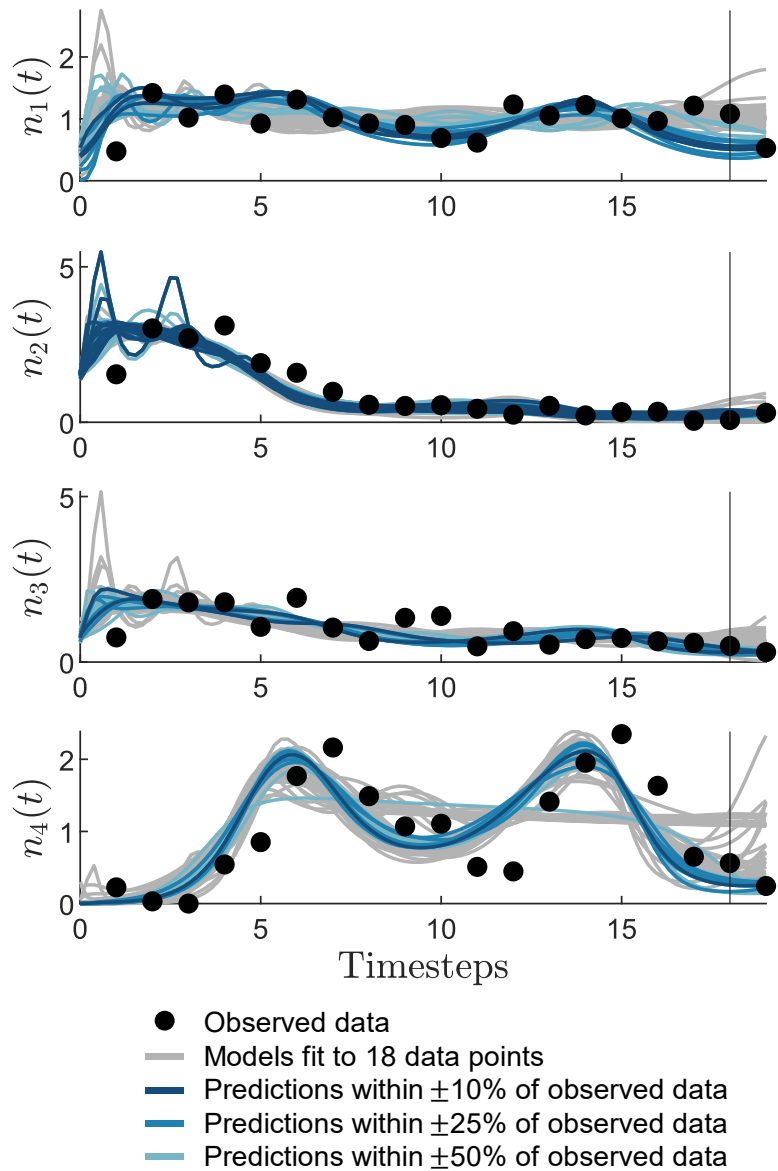

*Figure S2: The 50 best-fit Lokta-Volterra models when calibrated to the first 18 datapoints and used to predict the 19th. Timeseries predictions are coloured by the relative error of their predictions of the 19th datapoint.*

Across each of the species in each of the experiments, we assessed the percentage of models (of the 50 best-fit) which could predict within 10%, 25% and 50% accuracy of the observed datapoint (Figure S3). In all experiments, less than 50% of the models were able to predict within 10% of the true populations, and in most experiments, less than 10 models could forecast populations with accuracy (Figure S3, right). Only when our definition of "accurate predictions" slumped to a prediction within ±50% of the truth could the ecosystem models deliver even vaguely reliable results.

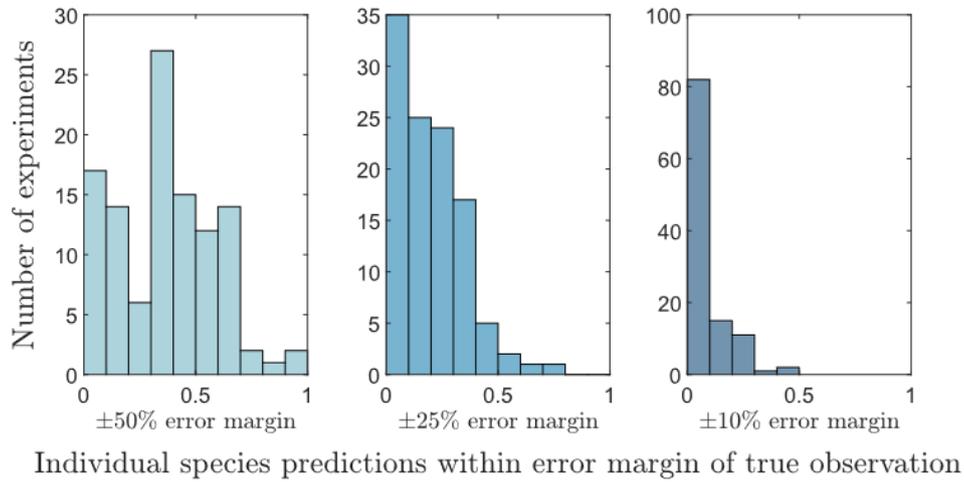

*Figure S3: The percentage of models (of 50 best-fit) that were able to forecast a single species' population 1 time-step ahead with a relative error less than 50%, 25% and 10% respectively.*

However, these results assume that managers are only interested in the predictions for a single species. When we calculate the percentage of best-fit models that can simultaneously predict *all* populations within some relative error, the results are very poor. Figure S4 reveals that the ecosystem models can effectively never predict all species populations accurately, even only 1 time-step ahead, even when only asking for predictions to be within ±50% of the truth.

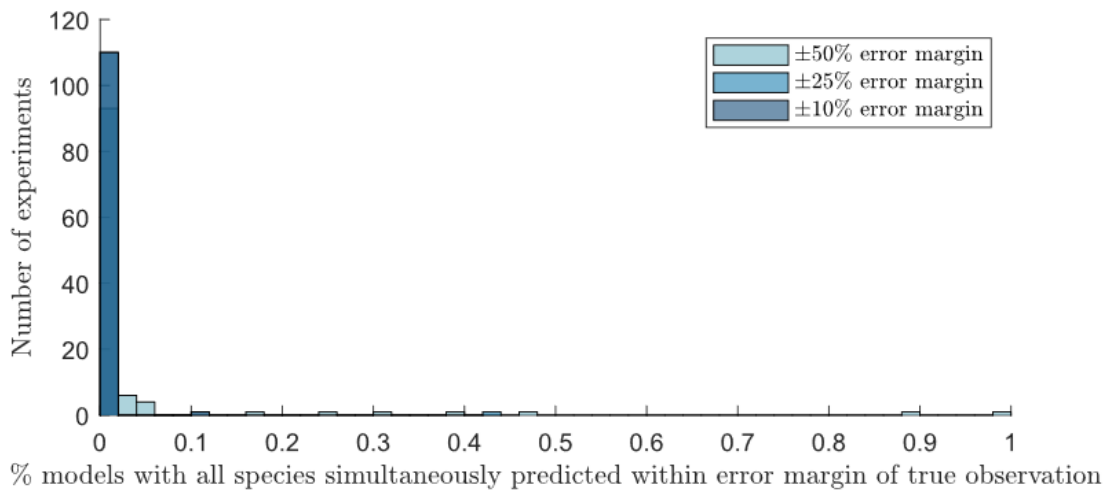

*Figure S4: The percentage of models (of 50 best-fit) that were able to forecast <u>all</u> species' populations 1 time-step ahead with a relative error less than 50%, 25% and 10% respectively.*

*2.2 Predictions compared to a simple model*

We then considered how the fitted ecosystem models perform in comparison to a simple null model. We choose our null model to be the zero-order approximation to any model, where all populations remain the same (i.e., that $n_i^m(t = 19) = n_i^o(t = 18)$). For each species $i$ in each experiment, we predicted the population at the 19th time point and calculated the relative error for both our best-fit ecosystem model, and this null model. Figure S5 shows that even though the null model is far simpler, it performs as well as the ecosystem models, with a slightly lower median error, but a lower error variance.

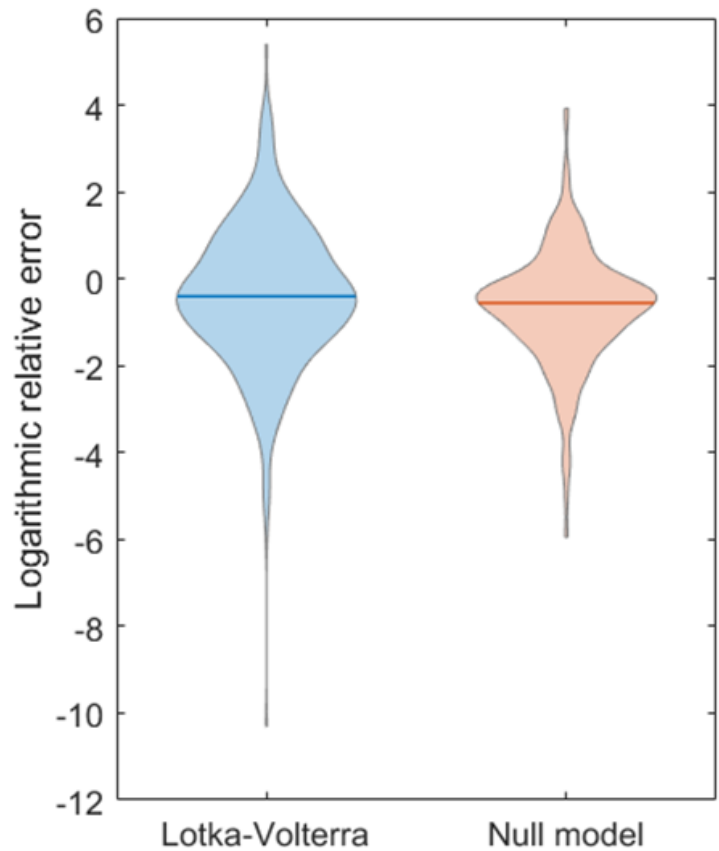

*Figure S5: The relative error of predictions of each species for each experiment using both the ecosystem models and the null models.*